\documentclass[a4paper,twoside,12pt]{article}
\makeatletter
\renewcommand{\fnum@figure}{\normalfont{\textbf{Fig.}} \thefigure}
\renewcommand{\fnum@table}{\normalfont{\textbf{Table}} \thetable}
\makeatother
\usepackage{graphicx}
\usepackage{titlesec}
\titlelabel{\thetitle.\;\;}
\titleformat*{\section}{\large\bfseries}
\titleformat*{\subsection}{\normalsize\bfseries}
\usepackage{amsthm}
\usepackage{natbib}
\usepackage{float}
\usepackage{geometry}
\usepackage{tikz}
\usepackage{eso-pic}
\usepackage{pdfpages}
\usepackage{afterpage}
\usepackage{setspace}
\usepackage[utf8]{inputenc}
\usepackage{pgfpages}
\usepackage{comment}
\usepackage{amssymb}
\usepackage{caption}
\usepackage{amsmath}
\usepackage{float}
\definecolor{ao}{rgb}{0.0, 0.5, 0.0}
\definecolor{grr}{rgb}{0.0, 0.5, 0.0}
\usepackage{hyperref}
\definecolor{dblue}{rgb}{0.0, 0.0, 0.55}
\hypersetup{colorlinks,
linkcolor={dblue},
citecolor={dblue},
urlcolor={violet}}  
\usepackage{enumitem}
\usepackage{xcolor}
\usepackage{listings}
\definecolor{coolblack}{rgb}{0.0, 0.18, 0.39}
\usepackage{textcomp}
\usepackage{amsmath,amssymb}
\usepackage{graphicx}
\DeclareUnicodeCharacter{0307}{$\ddot{o}$}
\usepackage{graphics}
\usepackage{color}
\usepackage{xcolor}
\usepackage[french,english]{babel}
\usepackage{systeme}
\usepackage{listings}
\usepackage{color} 
\definecolor{mygreen}{RGB}{28,172,0} 
\definecolor{mylilas}{RGB}{170,55,241}
\usepackage{float}

\begin{document}

\centerline{}

\centerline{}

\centerline {\Large{\bf A review of commonly used}}

\vspace{0.2cm}

\centerline{\Large{\bf compartmental models in epidemiology}}

\centerline{} \centerline{}

\newcommand{\mvec}[1]{\mbox{\bfseries\itshape #1}}

\centerline{{{Mohamed Mehdaoui} \footnote[1]{E-mail: mehdaouisimed@gmail.com}}}
\centerline{\emph{Faculty of Sciences and Technics of Settat, B.P. 577, Settat 26000, Morocco}} 

\centerline{}

\newtheorem{Theorem}{\quad Theorem}[section]

\newtheorem{Definition}[Theorem]{\quad Definition}

\newtheorem{Corollary}[Theorem]{\quad Corollary}

\newtheorem{Lemma}[Theorem]{\quad Lemma}

\newtheorem{rem}[Theorem]{\quad Remarks}

\centerline{\bf Abstract}\vspace{0.2cm} {\ In order to model an epidemic, different approaches can be adopted. Mainly, the deterministic approach and the stochastic one. Recently, a large amount of literature has been published using the two approaches. The aim of this  paper is to illustrate the usual framework used for commonly adopted compartmental models in epidemiology and introduce variant analytic and numerical tools that interfere on each one of those models, as well as the general related types of existing, ongoing and future possible contributions.} \\

{\vspace{0.8cm}  \emph{Keywords}:}  Dynamical systems; Epidemiological modeling;  Differential equations; Optimal control; Stability theory.

\newpage
\section{Introduction}

\hspace{0.3cm} Epidemiology is a scientific discipline that studies the distribution (who, when, and where) patterns and determinants of health and disease conditions in defined populations. It is a cornerstone of public health, and shapes policy decisions and evidence-based practice by identifying risk factors for disease and targets for preventive healthcare. The mathematical modeling of an epidemic allows to describe the spread of an infectious disease and predicts its evolution. This knowledge is  crucial to determine politics and measures that can control the fast spread of the disease. Mathematical models allow to estimate the effects of the epidemic, for instance the number of total people that are infected and the duration of the epidemic, moreover, the effects of prevention measures such as social distancing, vaccination or confinement. The study of the dynamics of an epidemic requires the knowledge of many structural variables: the absolute time, the age of the hosts, the duration of the infection, the immunity status. etc. Time is essential because it describes the continuous variation of a type of systems called dynamical systems. In the simple case, the system is formalized in terms of ordinary differential equations (ODEs). In some cases, to illustrate the "memory" effect of the dynamics, the system is formalized by fractional differential equations (FDEs), while delay differential equations (DDEs) interfere to capture the incubation period for instance. As for the partial differential equations (PDEs), they interfere when the spatial factor is included into the model. Deterministic dynamical systems, whether they're based on differential equations or partial differential equations, are easy to simulate. Their smoothness allows to explore a diversity of scenarios. Moreover, many theoretical and numerical tools allow to exploit these systems in a way to obtain a formula that expresses  the reproduction number in terms of the model's parameters. However, because the spread of a disease is a random process, sometimes It's pertinent to write the model in a probabilistic point of view. Moreover, the deterministic epidemiological models are formed under the assumption of a large-sized population. When it's not the case, the interactions between individuals are not uniform but have a random effect. In this case, the model is governed by stochastic differential equations (SDEs). \\

\hspace{0.3cm} Before choosing one of the two approaches of epidemiological modeling, the restriction to compartmental models is usually done. The logic behind this type of models is to express the dynamics of the population by dividing this latter into as many compartments as the clinical states. Usually, these compartments correspond to the population of susceptible (S), those who are infected (I) and those who don't participate to the epidemic anymore (R for recovered or removed individuals who are either dead or immune for a period of time). Variant considerations can be added to make the model more realistic and complex (exposed individuals but not yet infectious, infectious individuals but not yet detected, etc.). The transition from a clinical state to another is described by an incidence function. 
\\\\
 \begin{figure}[h!]
 \begin{center}
 \includegraphics[width=9.5cm]{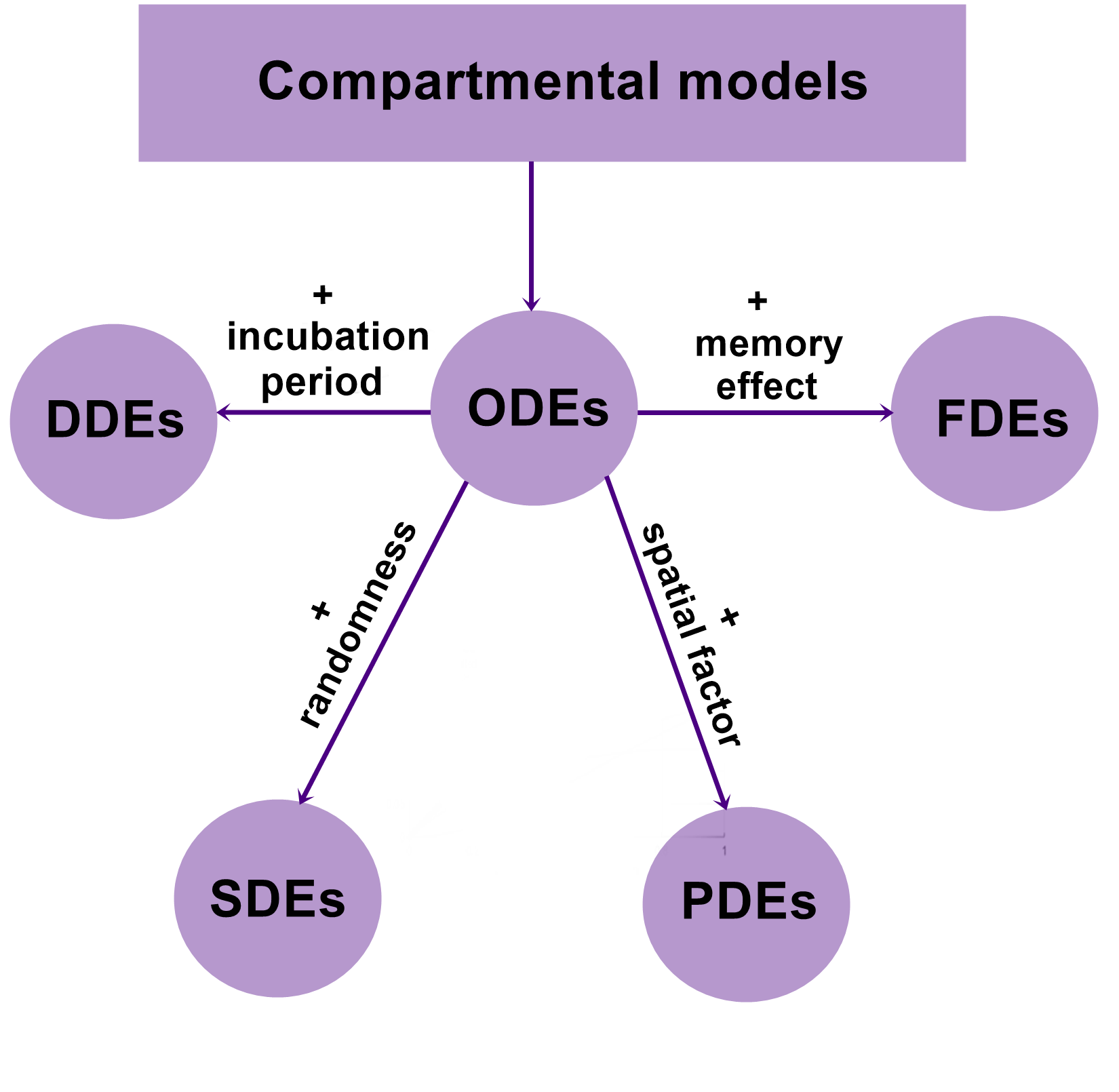}\\ 
 \caption{The relation between different approaches of compartmental modeling}
 \label{fig:fig}
 \end{center}
\end{figure} 
\newpage
\section{Common biological/mathematical background}
\hspace*{0.3cm} The compartmental mathematical models reviewed in this paper are expressed in either of the following forms\\
\begin{minipage}{0.47\linewidth}
\begin{equation}\label{eq:ode} 
\begin{cases}
\dfrac{du(t)}{dt}=f(u(t)) \; t>0,\\
\text{+ positive initial conditions}.
\end{cases}
\end{equation}
\end{minipage}
\hfill
\begin{minipage}{0.47\linewidth}
\begin{equation}\label{eq:sde}
\begin{cases}
du(t)=f(u(t)) dt+ A(u(t))\;dW(t) \; t>0,\\
\text{+ positive initial conditions}.
\end{cases}
\end{equation}
\end{minipage}
\hfill
\\
\begin{minipage}{0.47\linewidth}
\begin{equation}\label{eq:fde}
\begin{cases}
D^{\alpha} u(t)=f(u(t)) \; t>0,\\
\text{+ positive initial conditions}.
\end{cases}\;\;\;\;\;
\end{equation}
\end{minipage}
\hfill
\begin{minipage}{0.47\linewidth}
\begin{equation}\label{eq:dde1}
\begin{cases}
\dfrac{du(t)}{dt}=f(u_1(t-\tau_1),\cdots,u_m(t-\tau_m))\;\; t>0\\
\text{+ positive initial functions}.
\end{cases}\;\;\;\;\;
\end{equation}
\end{minipage}
\hfill
\begin{equation}\label{eq:pde}
\begin{cases}
-\underbrace{\mathcal{D} \Delta u(x,t)}_{\text{pointwise product}} + \dfrac{\partial u(x,t)}{\partial t}=f(u(x,t)) \; x \in \Omega, t>0,\\
\partial_\eta u_1(x,t)=\cdots=\partial_\eta u_m(x,t)=0 \; x \in \partial \Omega, t>0,\\
\text{+ positive initial conditions}.
\end{cases}
\end{equation}
 such that
 \begin{itemize}
 \item $u:=(u_1,\cdots,u_m)$ with $m\in \mathbb{N}^*$ representing the number of compartments.
 \item $f: (0,\infty)^m \longrightarrow \mathbb{R}^m$ is a continuously differentiable function.
 \item $\dfrac{d}{dt}$   is the classical time derivative which will be denoted by $'$ for  $\eqref{eq:ode}$ and $\eqref{eq:dde1}.$ 
 \item $ \dfrac{\partial }{\partial t}$ is the partial derivative with respect to time.
 \item $\Omega \subset \mathbb{R}^p$ ($p \in \{1,2\}$) is an open bounded set, $\partial \Omega$ its boundary and $\partial _\eta$ is the outward normal derivative.
 \item $\Delta u:=(\Delta u_1,\cdots,\Delta u_m)$ where $\Delta:=\dfrac{\partial^2 }{\partial x_{1}^2}+\dfrac{\partial^2 }{\partial x_{2}^2}$ is the Laplace operator.\\
 \item $\mathcal{D}:=(d_1,\cdots,d_m)$ is a vector of strictly positive diffusion coefficients. 
  \item $W_t$ is an $n$-dimensional  Wiener process (will also be denoted by $B_t$).
 \item $A(u(t)) \in \mathcal{M}_{m \times n}(\mathbb{R})$ is a $m \times n$ matrix with locally Lipschitz coefficients.
 \item $D^{\alpha}$ is some fractional time derivative of order $\alpha \in (0,1)$   (see \cite{review}).
 \item $\tau_i \geq 0 \; \forall i \in \{1,\cdots,m\}$ are the discrete time delays.
 \end{itemize}

For more terminology on each of the previous differential equations, we refer to \cite{3,4,5,pd}.\\
All the considered compartmental models  share a common biological/mathematical framework as well as some steps followed in their mathematical and qualitative analysis such as:
\paragraph{The basic reproduction number $\mathcal{R}_0$:} 
One quantity that plays a crucial role on all of the approaches  of epidemiological modeling illustrated in Figure $\ref{fig:fig}$, is the basic reproduction number, which is biologically speaking, defined as  the average number of secondary infected caused by a single infected individual on the period of his contagiousness. One question that intuitively arises is: \textit{How this quantity can be expressed mathematically based on the chosen compartmental model?} The answer was given by  \cite{1}. The problem was reduced to a computation of the spectral radius of the product of two matrices  $F$ and  $V^{-1}$, such that $F$ represents the rate of appearance of new infections and $V$ is the rate of transfer of individuals into infected compartments by all other means.
\paragraph{The study of mathematical and biological well-posedness:} The mathematical well-posedeness of a compartmental model relies  on proving the existence and uniqueness of the solution. Depending on the type of the differential equation that governs the model, there exist many theorems that can be applied to achieve this purpose. For instance, the Cauchy-Lipschitz theorem for \eqref{eq:ode}  (see \cite{3} pp. 1-8). For \eqref{eq:sde}, this is established by means of  the Itô existence theorem  (see \cite{4} pp. 65-84). For \eqref{eq:dde1},  (see \cite{5} pp. 13-23). Finally for \eqref{eq:pde},  (see \cite{2} pp. 249-261). Once the mathematical well-posedeness is achieved, the biological well-posedness relies on proving the boundedness of the unique solution as well as its positiveness. Given a positive initial condition, for $\eqref{eq:ode},\eqref{eq:sde},\eqref{eq:fde}$ and $\eqref{eq:dde1},$ this is usually proven by a contradiction argument. For $\eqref{eq:pde}$, the maximum principle approach is adopted  (see \cite{max}).
\paragraph{The study of different disease equilibria:}  In dynamical systems, it is of great importance to study the solutions that don't change in time. Such solutions are called equilibria. For compartmental models, two equilibria are crucial: the \underline{Free Disease Equilibrium (FDE)} and the \underline{Endemic Equilibrium (EE)}. The first equilibrium, is obtained  by assigning the value zero to all infected classes, and then solving the equation $f(u)=0,$ to deduce the rest of the terms of the FDE. On the other hand, for the EE, under the hypothesis $\mathcal{R}_0>1.$ Solving the system $f(u)=0$ yields the other corresponding equilibrium. We emphasize that for $\eqref{eq:ode}, \eqref{eq:pde}, \eqref{eq:fde}$ and $\eqref{eq:sde}$ the EE  can be obtained by simple or long algebraic manipulations depending on the number of compartments, whereas for $\eqref{eq:dde1},$  even for a small number of compartments, further analysis is required. Once the existence of the FDE and EE equilibria is shown, their local and global stability must be studied. For most of the previous models, the corresponding system is linearized around the corresponding equilibrium. For the FDE (resp EE), under the hypothesis $\mathcal{R}_0<1,$ (resp $\mathcal{R}_0>1$) it can be shown that the characteristic equation associated to the linearized system has roots with strictly negative real part which yields the local asymptotic stability of the FDE (resp EE). One should know that the conditions under which the roots have strictly negative real parts are not always easy to obtain, especially when the number of compartments increases. But, In virtue of the Routh-Hurwitz criterion (see \cite{routh}), assumptions on the parameters can be concluded.  On the other hand, the global stability can be very challenging for all the previous models, given that the latter requires a suitable choice of the Lyapunov function and unfortunately, no mathematical method can be used to get the suitable choice, but several authors have tackled the problem and proposed a general form that can work as Lyapunov functions for some particular epidemiological models. For instance, \cite{lyapart1,lyapart2,hattaflyap}. If the Lyapunov function is found, under the same assumptions on $\mathcal{R}_0,$ and sometimes further assumptions depending on the model, the global stability of the equilibria can be proved using the classical Lyapunov stability theory   (see \cite{stab}).
\paragraph{Numerical simulation of the model:} Once the mathematical analysis is concluded, to support the theoretical results, a numerical simulation is in order. Before the simulation, the question of the values assigned to the parameters arises. One approach to solve this problem is to consider an estimation of the parameters, but such an approach requires availability of the data up to the present within  governments' institutions. As an example, in the case of models taking the form \eqref{eq:ode}, we refer to \cite{iv}, where the authors used the method of variational imbedding (MVI) to identify the transmission rate $\beta>0$ as well as the recovery rate $\gamma>0$ of the SIR model. We highly recommend    \cite{6} to the reader to get familiar with the approach used in the cited paper. A different approach was used by \cite{7}, the authors developed a numerical formula estimating the parameters of the SIR model, based on the approximation of the classical derivative (first order and second order) of the model and the minimization of a least square sum. Once the parameters are estimated, the numerical simulation is carried on and a variety of numerical methods can be used
for this purpose. For instance, for \eqref{eq:ode}, the fourth order Runge-Kutta  method is preferred due to its high accuracy. for \eqref{eq:fde}, the Adams-Bashforth method or its generalized version can be used. For \eqref{eq:pde}, one can think of the Finite Difference Method or the Operator Splitting Method. For \eqref{eq:dde1}, a modified version of Runge-Kutta methods can be adopted. Finally, for \eqref{eq:sde}, one can use the Euler–Maruyama method or the Milstein's higher order method. To get a better understanding of these methods we refer to \cite{9,10,11,111,12,sto,opsp}.
\paragraph{The study of the optimal control problem:} When the model is simulated numerically, particularly, for the case $\mathcal{R}_0 >1,$ the problem of finding control strategies that stop the epidemic from spreading further is studied. Usually this control measures reside on introducing treatment and vaccination controls to the studied model. The study of the optimal control problem is carried on theoretically at first by proving the existence and uniqueness of such an optimal control, and then by using direct or indirect numerical methods to compute it. For \eqref{eq:ode}, \eqref{eq:sde} \eqref{eq:fde} and \eqref{eq:dde1}, usually the indirect method is preferred due the reduction of the problem to a system of initial value problems in virtue of the Pontryagin maximum principle and some of its variations, this system is then solved by means of the Forward-Backward Sweep Method. We refer to \cite{opde} to gain more knowledge on both theoretical and numerical aspects of optimal control theory and  \cite{oc} for its application to biological models.
\section{Review of some of the main contributions in compartmental models}
\subsection{The foundation}
\hspace{0.3cm} Strictly speaking, the birth of compartmental epidemiological models was attributed \cite{kerm}. They formalized the concept of compartmental models by using a set of ordinary differential equations to describe the behavior of an epidemic. Their method is still considered  valid, and is largely used in  recent research. The main objective of the two researchers was to understand the reasons why the pandemic of Spanish Flu didn't infect the whole population. The model is composed into three compartments as the following figure shows
 \begin{center}
 \begin{figure}[h!]
 \includegraphics[width=13cm]{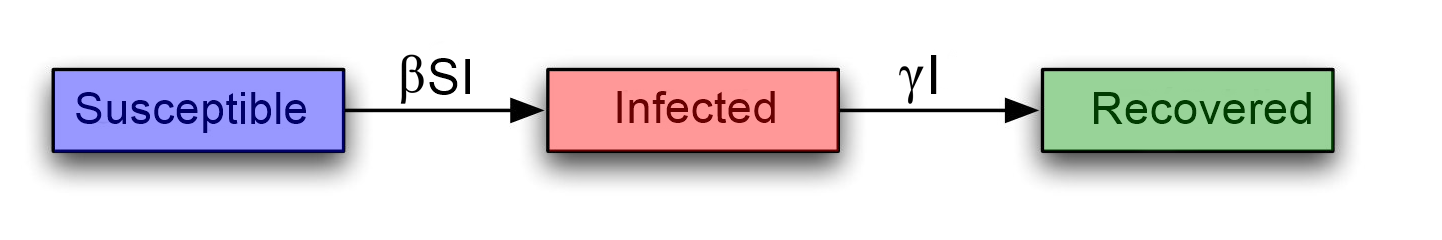}\\ 
 \caption{Diagram of the SIR model.}
 \label{fig:sirfig}
\end{figure} 
\end{center}
\hspace*{0.3cm} The three compartments are considered as follows \\
\begin{description}
\item[$\bullet$] $S(t):$ the class of those individuals who are capable of contracting the disease and becoming infected. 
\item[$\bullet$] $I(t):$ the class of infected individuals, those who
capable of transmitting the disease to others.
\item[$\bullet$] $R(t):$ the class of those individuals who have had
the disease and are dead, or have recovered and are permanently
immune or are isolated. 
\end{description}
This model has two parameters defined by
\begin{description}
\item[$\bullet$] $\beta >0:$ the infection/transmission rate.  
\item[$\bullet$] $\gamma >0:$ the recovery/removal rate.
\end{description}
The differential system governing the original SIR model is:
\begin{equation}\label{eq:sir}
\begin{cases}
S^{\prime}(t)=-\beta I(t) S(t)\;\; t>0, \\
I^{\prime}(t)=\beta I(t) S(t)-\gamma I(t) \;\; t>0, \\
R^{\prime}(t)=\gamma I(t) \;\; t>0, \\
\text{+ Initial positive conditions.}
\end{cases}
\end{equation}
such that $N=S(t)+I(t)+R(t) \;\;\; \forall t\geq 0$ is the total population. \\ 

\hspace{0.3cm} It is important to state that the original SIR model has some limitations. For instance, the vital dynamics (natural birth/death rate) are not captured by the model, also, the model assumes that all the infected people are infectious and are spreading the disease among the susceptible population, moreover, it doesn't consider the incubation period. Consequently,
the diseases for which this model can be applied are limited and further contributions to make it more practicable are required.\\\\
 \textbf{\underline{Notation:}} All the   reviewed models' parameters will be renamed if necessary to unify the notation. Unless stated otherwise, $\Lambda, \beta, \gamma ,\mu , d, \mathcal{R}_0$ and the letters S,I,R,E,V,D,A,H,Q,C,T and M hold the same definitions that will be stated thereafter.
\subsection{Some of the main contributions to  \eqref{eq:ode}}

\hspace{0.3cm} The contributions of this form take into account the following
\begin{itemize}
\item \textbf{\underline{Extension of the SIR model:}} This is usually done by considering more clinical states resulting in the addition of more compartments to the model. Examples of such added compartments are the exposed (E), the vaccinated (V), the deceased (D), the asymptomatic (A), the hospitalized (H), the quarantined (Q), the  cross-immune (C), the treated (T), the maternally-derived immune (M), etc. To each interaction of a compartment with another, a positive parameter is added. Furthermore, to make the modified models more realistic,  further vital parameters (natural birth rate $\lambda>0$, natural mortality rate $\mu>0$, death rate caused by the disease $d>0$) are taken into account. Another type of extension concerns diseases in which different species may intervene (ex: Humans and Mosquitoes for Dengue fever), or while considering heterogeneous host populations. In this case, multi-group SIR and their extensions are considered. 
\item \textbf{\underline{Modification/Generalization of the bilinear incidence rate:}} It is known that the bilinear incidence rate $"\beta SI"$ of the original SIR model is based on the law of the chemical mass action stating that the variation of the rate of a chemical reaction is directly proportional to the product of the activities or concentrations of the reactants. This incident rate is classic, and does not take into account the saturation phenomena for a large number of infected individuals. One way to remedy this problem is to consider other rates defined by $g(I)S,$  $g(I)h(S),$ or in a more general way, one of the form $f(S,I),$ where $f,h$ and $g$ are functions satisfying adequate assumptions  \cite{gkm,gkm2}. Some of the commonly  well-known incidence rates since the $20^{th}$ century are:
\begin{itemize}
\item[$\bullet$] $\dfrac{\beta S I}{N}:$ Standard Incidence rate \cite{std}.
\item[$\bullet$] $\dfrac{\beta S I}{1+a I}:$\;\; Holling type incidence rate \cite{may}.
\item[$\bullet$] $\dfrac{\beta S I}{1+aS+bI}:$\;\; Beddington-DeAngelis incidence rate \cite{bed}.
\item[$\bullet$] $\dfrac{\beta S I}{(1+aS)(1+\gamma I)}:$\;\; Crowley-Martin incidence rate \cite{crowley}.

\end{itemize}
Such that $a$ and $b$ are strictly positive constants, $N$ is the sum of all the considered compartments, while $\beta$ and $\gamma$ are respectively the transmission and the recovery rates. Note that for a fixed value of $S,$ and for a very large value of the infected class, all the above incidence rates become bounded and the crowding effect is taken into account which is practically acceptable. Not the same can be said in the case for the classical bilinear rate. 
Below, we present a figure showing the four incidence rates along with the classical incident rate for the SIR model.
 \begin{figure}[H]
\begin{center}
 \includegraphics[width=9.7cm]{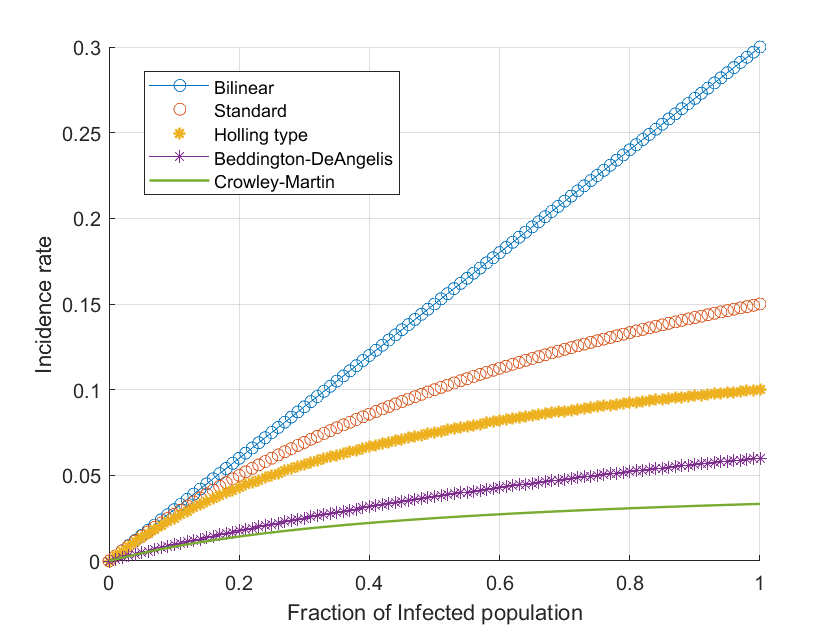}\\ 
 \caption{Some of the early proposed incidence rates compared to the bilinear incidence rate}
 \label{fig:sirfig}
 \end{center}
\end{figure} 
\hspace*{0.3cm} The work on new proposed incidence rates and extended SIR models continued in the $21^{st}$ century. Indeed, \cite{xiao} worked with the incidence rate given by  $\dfrac{\beta I S}{1+\alpha I^{2}}$ with $\alpha>0.$ Their aim was to take into account the change on the behavior of the susceptible population when they become aware of the high number of infected individuals. Their new proposed SIRS model was expressed in the following system of odes 
\begin{equation}
\begin{cases}
\begin{aligned}
&S'(t)=\Lambda-\mathrm{\mu} S(t)-\frac{\beta S(t) I(t)}{1+\alpha I^{2}(t)}+\kappa R(t), \\
&I'(t)=\frac{\beta S(t) I(t)}{1+\alpha I^{2}(t)}-(\mu+\kappa) I(t), \\
&R'(t)=\gamma I(t)-(\mu+\kappa) R(t),\\
&\text{+ positive initial conditions.}
\end{aligned}
\end{cases}
\end{equation}
with one added parameter $\kappa>0,$ being the rate in which the recovered lose immunity and return to the susceptible class. \\

\hspace{0.3cm} In the same year, the research related to incidence rates begun to steer to the generalization of the previous proposed ones. In fact, \cite{korob} proposed the following SIR model with a general incidence rate $f(S,I)$\\
\begin{equation}
\begin{cases}
\begin{aligned}
&S'(t)=\mu-f(S(t), I(t))-\mu S(t), \\
&I'(t)=f(S(t), I(t))-(\mu+\gamma) I(t), \\
&R'(t)=\gamma I(t)-\mu R(t),\\
&\text{+ positive initial conditions.}
\end{aligned}
\end{cases}
\end{equation}
where the birth/death rates are taken to be equal to $\mu>0$ and such that $f$ is is a continuously differentiable function satisfying the following assumptions
\begin{enumerate}
\item $f:(S, I) \in Q \rightarrow f(S, I) \in \mathbb{R}^{+}$ is a monotonically increasing function \; $\forall S,I>0.$ 
\item $f(S, 0)=f(0, I)=0 \; \forall S,I>0.$ 
\item $f$ is concave with respect to I.
\end{enumerate}
\centerline{  where $Q:=[0,1] \times[0,1] \subset \mathbb{R}_{+}^{2}.$}
\begin{rem}
It should be noted that while the assumptions are biologically intuitive, each one serves a mathematical purpose. Assumption 1 is for the mathematical well posedness, while 2 is for the existence of the FDE and 3 is for the global stability analysis of the EE. 
\end{rem}
\hspace*{0.3cm}  \cite{korob} proved the global assymptotic stability of the FDE (resp EE) under the condition $\mathcal{R}_0:=\dfrac{1}{\mu+\gamma} \dfrac{\partial f(S,I)}{\partial I}\bigg|_{FDE}\leq1$ (resp  $\mathcal{R}_0>1$)  by means of the Lyapunov function. Later, \cite{buonomo} considered the same  SIRS model with general incidence rate, and they proved the local stability of the FDE and the EE under weaker assumptions on the general incidence rate $f(S,I).$\\
The work on the development of adequate incidence rates for the SIRS model continued. For instance, \cite{li} considered a general incidence rate of the form $f(I)S$ with $f$ verifying some adjusted assumptions.
\item[$\bullet$]\textbf{\underline{Introduction of immigration:}} Parallel progress in ameliorating the SIR model from a different perspective begun when \cite{mccluskeyim} proposed an SEI model with immigration. Their aim was to take into account individuals that enter a population and their contribution into maintaining the presence of tuberculosis disease. The associated model was expressed in the following system of odes 
\begin{equation}
\begin{cases}
\begin{aligned}
&S'(t)=(1-p-q) \Lambda-c \beta \frac{S(t) I(t)}{N(t)}-\mu S(t)+r_{1} E(t)+r_{2} I(t), \\
&E'(t)=p \Lambda+c \beta \frac{S(t) I(t)}{N(t)}-\left(k+\mu+r_{1}\right) E(t), \\
&I'(t)=q \Lambda+k E(t)-\left(\mu+d+r_{2}\right) I(t),\\
&\text{+ positive initial conditions.}
\end{aligned}
\end{cases}
\end{equation}
with the following assumptions 
\begin{itemize}
\item[$\bullet$] $c>0$, $\Lambda$ is the total rate in which new individuals enter the whole population while $(1-p-q) \Lambda$ in the rate in which they enter the susceptible class, $p \Lambda$ the one in which they enter the exposed class and finally $q \Lambda,$ the one is which they enter the infected class. With the assumption $p+q \in [0,1].$ 
\item[$\bullet$] $r_1$ (resp $r_2$) is a strictly positive rate in which exposed (resp infected) are treated . 
\end{itemize}
\hspace*{0.3cm} In the same paper, \cite{mccluskeyim}  tried to generalize the recruitment rate of immigrants by considering the recruitment function $B(N),$ where $N:=S+E+I.$ The following assumptions were made 
\begin{itemize}
\item[$\bullet$](H1) There exists a unique $N_{0}>0$ such that $B\left(N_{0}\right)-\mu N_{0}=0$.
\item[$\bullet$] (H2) $B^{\prime}(N)<\frac{B(N)}{N}$ for $N \in\left(0, N_{0}\right]$.
\item[$\bullet$] (H3) $b_{0}=\lim _{N \rightarrow 0+} \frac{B(N)}{N}>\mu$, allowing that $b_{0}$ may be infinite. 
\end{itemize}
$(H1)$ and $(H3)$ are biologically related assumptions while $(H2)$ is to ensure the existence and uniqueness of the endemic equilibrium. \\
Later on, several authors' aim was to extend the previous models into ones that take into account the concept of immigration. 
\\\\
For instance, \cite{cui} considered an SIRS model with a constant immigration and an incidence rate of the form $g(I)S$ with $g$ verifying two assumptions. \\\\
Driven by the same motivation, \cite{khan} considered an SEI model with immigration and a general incidence rate of the form $f(S,I)$. 
\item[$\bullet$]\textbf{\underline{Introduction of the treatment:}} One last angle of contribution that we discuss in this section is the one related to the finding of a suitable treatment function. This begun when \cite{wendi} remarked that the removal rate being linear in terms of the infected in not practicable because of the limited capacity of treatment in each country. This resulted in the proposal of the following treatment function 
$$h(I)= \begin{cases}r & \text { for } I>0, \\ 0 & \text { for } I=0.\end{cases}$$ 
where $r > 0$ is a constant representing the capacity of treatment for infectives.\\
The model is formulated in the following system of odes 
\begin{equation}
\begin{cases} 
\begin{aligned}
&S'(t)=\Lambda-\mu S(t)-\beta S(t) I(t), \\
&I'(t)=\beta S(t) I(t)-(\mu+\gamma) I(t)-h(I(t)), \\
&R'(t)=\gamma I(t)+h(I(t))-\mu R(t),\\
&\text{+ positive initial conditions.}
\end{aligned}
\end{cases}
\end{equation}
\hspace*{0.3cm} One limitation of the proposed treatment function is that a constant treatment is always considered even when the treatment capacity is not reached.\\ \cite{wang} solved this limitation by considering the following treatment function 
$$
T(I)= \begin{cases}r I & \text { if } 0 \leqslant I \leqslant I_{0}, \\ k & \text { if } I>I_{0}.\end{cases}
$$
where $k=r I_0.$ The modified model was given by 
$$
\begin{cases} 
\begin{aligned}
&S'(t)=\Lambda-\mu S(t)-\beta S(t) I(t), \\
&I'(t)=\beta S(t) I(t)-(\mu+\gamma+d) I(t)-T(I(t)), \\
&R'(t)=\gamma I(t)+T(I(t))-\mu R(t),\\
&\text{+ positive initial conditions.}
\end{aligned}
\end{cases}
$$
\hspace*{0.3cm} For an outbreak disease such as  SARS,  \cite{zhonghua} thought replacing the treatment rate function $T$ with a saturated rate of Holling type would be more pertinent. The considered treatment function was
$$
h(I)=\frac{\beta I}{1+\alpha I},\;\;\; \alpha>0.
$$
\hspace*{0.3cm}  
\vspace*{0.4cm}\\
The related contributions done after were mostly proposing new models obtained by taking combinations of saturated rates for the treatment function and some of the recent studied incidence rates introduced in this paper. \\\\ 
We give a figure illustrating the reviewed treatment rates as well as some that have been inspired from the well-known incidence rates. 
 \begin{figure}[h!]
 \begin{center}
 \includegraphics[width=9cm]{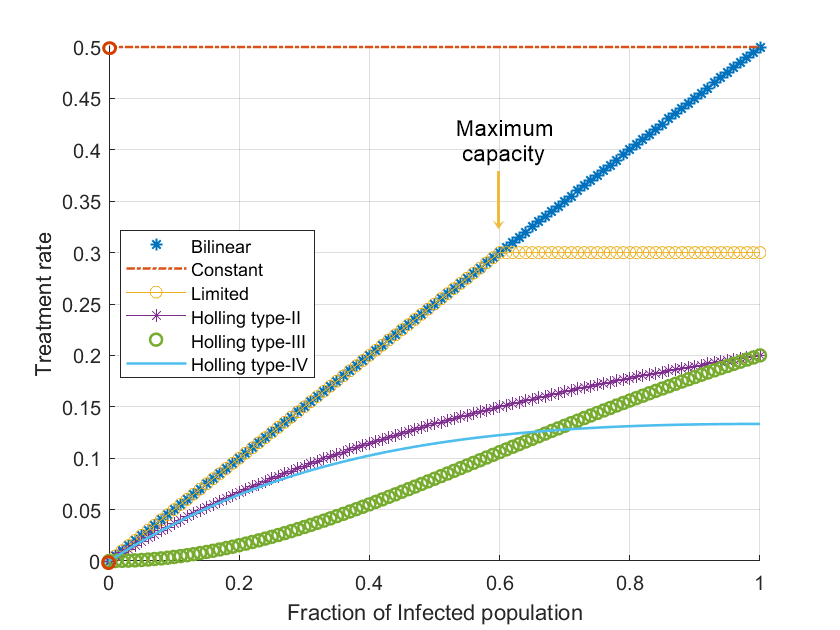}\\ 
 \caption{Some of the commonly used treatment rates}
 \label{fig:sirfig}
 \end{center}
\end{figure}
\end{itemize}
\hspace*{0.3cm} Below, we present a table summarizing the previous discussed contributions falling into  $\eqref{eq:ode}$ 
\begin{table}[H]
\begin{center}
\caption{Some of the main contributions falling into category $\eqref{eq:ode}$}
\begin{tabular}{|p{3cm}|l|l|p{2.5cm}|p{5cm}|}
\hline
Ref. & Model &  Incidence & Treatment & Contributions \\ \hline
\cite{xiao} & SIR & $\dfrac{\beta IS}{1+\alpha I^2},\;\; \alpha>0$ & None & Proposing  a new incidence rate taking into account the behavioral change towards the infected. \\ \hline
\cite{korob} & SIR & $f(S,I)$ & None & Proposing a general incidence rate for the SIR  model, global stability of the  DFE \& EE for the model under some assumptions on $f.$ \\ \hline
\cite{buonomo} & SIRS & $f(S,I)$ & None & Local stability of the SIRS model with general incidence rate under weakened conditions on $f.$ \\ \hline
\end{tabular}
\end{center}
\end{table}
\begin{table}
\begin{center}
\caption{Some of the main contributions falling into category $\eqref{eq:ode}$ continued}
\begin{tabular}{|p{3cm}|l|l|p{2.5cm}|p{5cm}|}
\hline 
Ref. & Model &  Incidence & Treatment & Contributions \\ \hline
\cite{mccluskeyim} & SEI & Bilinear & Constant treatment rate, General treatment rate & Proposition of a new SEI model taking into account the affect of immigration, generalization of the model by considering a general immigration rate,  global stability with constant/general immigration rate. \\ \hline
\cite{khan} & SEI & $f(S,I)$ & None &  	Proposing  a  SEI model with a general incidence rate and constant immigration rates within the the infected and the exposed, global stability of the FDE and EE, Numerical simulation for the particular bilinear incidence rate.  \\ \hline
\cite{wendi} & SIR & Bilinear & treatment rate with limited capacity & Proposing  an SIR model with a more realistic treatment rate taking into account the limited resources, bifurcation analysis, introducing conditions under which the disease is instinct and limited treatment is sufficient. \\ \hline
\cite{wang} & SIR & Bilinear & Infected-depending treatment rate with limited capacity & Tackling the limitation of the treatment rate in \cite{wendi} by introducing a rate proportional to the infected class as long as the capacity is not reached, study of existence and uniqueness of the DFE \& EE, bifurcation analysis. \\ \hline
\end{tabular}
\end{center}
\end{table}
\newpage
It should be noted that the efficiency of contributions falling into category \eqref{eq:ode} depends on the better understanding of the modeled disease. For a simple one, the limitation to a reduced number of compartments and a bilinear incidence rate may suffice  \cite{sirvit}. But for complicated diseases, such as Covid-19 or seasonal diseases, a large number of compartments is required for the first and an adequate choice of the incidence rate is needed for the second to capture the maximum of possible intervening factors in each one of the modeled diseases \cite{torr,svir3}.\\\\ \hspace*{0.3cm} As far as the contributions of category $\eqref{eq:ode}$ can go, this type of modeling will always have several limitations, some of which are:
\begin{enumerate}
\item[$\bullet$] \textbf{Limitation 1:}  They don't capture the uncertainty and variability that is inherent in real-life epidemics due to factors such as the unpredictability of person-to-person contact.
\item[$\bullet$] \textbf{Limitation 2:}  They don't take into account the memory effects on their dynamics (dependency on the past). As a matter of fact, given the knowledge of the history of a disease, people can use different precautions that affect the change in the dynamics. 
\item[$\bullet$] \textbf{Limitation 3:} They're based on the assumption of instant cause and effect which is not correct in real life. For instance, an infected individual needs some time-period before becoming infectious and starts spreading the disease. Furthermore, when treatment is applied, a time-period is also needed in order for the effect of the treatment to start showing.
\item[$\bullet$] \textbf{Limitation 4:}  They don't take into account the spatial factor. For infectious diseases, the movement of individuals is a major factor of the disease spread, thus the densities of different compartments of the model shouldn't be just in terms of time but  also space.
\end{enumerate} 
\subsection{Some of the main contributions to  \eqref{eq:sde}}

\hspace{0.3cm} The essential aim of contributions falling into category $\eqref{eq:sde}$ is to overcome "\textbf{Limitation 1}" by adapting all the previous contributions of category $\eqref{eq:ode}$ to a stochastic point of view. \\\\
\hspace*{0.3cm} \cite{tornatore} adjusted the deterministic SIR model to the following stochastic one 
\begin{equation}\label{eq:torn}
\begin{cases}
\begin{aligned}
&\mathrm{d} S(t) =(-\beta S(t) I(t)-\mu S(t)+\mu) \mathrm{d} t-\sigma S(t) I(t) \mathrm{d} W(t), \\
&\mathrm{d} I(t) =(\beta S(t) I(t)-(\gamma+\mu) I(t)) \mathrm{d} t+\sigma S(t) I(t) \mathrm{d} W(t), \\
&\mathrm{d} R(t) =(\gamma I(t)-\mu R(t)) \mathrm{d} t,\\
&\text{+ positive initial conditions.}
\end{aligned}
\end{cases}
\end{equation}
with the natural birth and death rates being equal to $\mu>0.$ The authors were able to prove that the global stability of the DFE (resp EE) holds under the condition $0<\beta<\min \left\{\gamma+\mu-\frac{\sigma^{2}}{2}, 2 \mu\right\}.$ (resp $ \beta>\lambda+\mu+\frac{\sigma^{2}}{2}$). One can remark that for $\sigma=0,$ this condition takes into account that $\mathcal{R}_0:=\dfrac{\beta}{\gamma+\mu}<1$ (resp $\mathcal{R}_0>1$) which is exactly the same condition that assures the asymptotic global stability of the DFE (resp EE) associated the deterministic counterpart of the considered stochastic SIR model. Numerical simulations showed important oscillations of the obtained solutions, which captures the random behavior of disease dynamics in reality.\\\\
\hspace*{0.3cm} \cite{lu} extended \eqref{eq:torn} by  taking into account the loss of immunity to the disease. The author proposed the following SIRS model 
\begin{equation}\label{eq:lu}
\begin{cases}
\begin{aligned}
&\mathrm{d} S(t)=(-\beta S(t) I(t)-\mu S(t)+\gamma R(t)+\mu) \mathrm{d} t-\sigma S(t) I(t) \mathrm{d} W(t), \\
&\mathrm{~d} I(t)=(\beta S(t) I(t)-(\gamma+\mu) I(t)) \mathrm{d} t+\sigma S(t) I(t) \mathrm{d} W(t), \\
&\mathrm{~d} R(t)=(\gamma I(t)-(\mu+\lambda) R(t)) \mathrm{d} t,\\
&\text{+ positive initial conditions.}
\end{aligned}
\end{cases}
\end{equation}
Under the assumption $\beta<\lambda+\mu-\frac{\sigma^{2}}{2}$, stochastic asymptotic stability of the FDE has been proved based on the Lyapunov method.\\\\
\hspace*{0.3cm} \cite{jyang} proposed the following SIR model with a different added noise term 
\begin{equation}
\begin{cases}
\begin{aligned}
&\mathrm{d} S(t)=(\Lambda-\beta S(t) I(t)-\mu S(t)) \mathrm{d} t+\sigma_{1} S(t) \mathrm{d} B_{1}(t), \\
&\mathrm{d} I(t)=(\beta S(t) I(t)-(\mu+d+\gamma) I(t)) \mathrm{d} t+\sigma_{2} I(t) \mathrm{d} B_{2}(t), \\
&\mathrm{d} R(t)=(\gamma I(t)-\mu R(t)) \mathrm{d} t+\sigma_{3} R(t) \mathrm{d} B_{3}(t),\\
&\text{+ positive initial conditions}.
\end{aligned}.
\end{cases}
\end{equation}
with natural birth rate (resp death rate) $\Lambda>0$ (resp $\mu>0$) and a death rate caused by the disease $d>0$ as well as different stochastic intensities $\sigma_i >0$\; $ i \in \{1,2,3\}$ and different Wiener processes $B_i$ \; $ i \in \{1,2,3\}.$ 
The authors proved the existence and uniqueness of a global positive solution to the model. Moreover, it was shown that if $\sigma_{1}^{2}<\mu$ and  $\sigma_{2}^{2}<2(\mu+d+\gamma)$  then the boundedness of solutions is assured, they also studied the asymptotic behavior of the solution around the deterministic FDE and EE equilibria associated the deterministic counterpart of the proposed model. Precisely, if the boundedness assumptions are verified and $\mathcal{R}_0:=\dfrac{\beta \Lambda}{\mu(\mu+d+\gamma)}<1$ then the oscillations of solutions around the DFE associated to the deterministic counterpart of the model increase in terms of $\sigma_1$ and $\sigma_2.$ For $\mathcal{R}_0,$ the oscillations also occur around the state $\left(\dfrac{2 \mu}{2 \mu-\sigma_{1}^{2}} S^{*}, \dfrac{2 \mu\left(\mu+d+\gamma-p \gamma^{2}\right)}{2 \mu(\mu+d+\gamma)-p \gamma^{2}-\mu \sigma_{2}^{2}} I^{*}, \dfrac{\mu}{\mu-\sigma_{3}^{2}} R^{*}\right)$ provided that $\sigma_{1}^{2}<2 \mu, \quad \sigma_{2}^{2}<2(\mu+d+\gamma), \quad \sigma_{3}^{2}<\mu$. Such that the EE is the endemic equilibrium of the deterministic counterpart of the model.  \\\\
The numerical simulations were done using the Milstein's Higher Order
Method \cite{sto} to  support the obtained theoretical results and showed that for a small value of the stochastic intensities, the boundedness of the solution is assured and for a large time, the different stochastic equilibria approach their deterministic counterparts.
\\\\
\hspace*{0.3cm} \cite{rao} modified $\eqref{eq:lu}$ by taking a Holling-type incidence rate, the noise was added to the three compartments and was supposed proportional to the distances of the states from the steady ones corresponding to the endemic equilibrium of the deterministic counterpart of the model. The proposed model was as follows 
\begin{equation}
\begin{cases}
\begin{aligned}
&\mathrm{d} S(t)=\left(\Lambda-d S(t)-\frac{\beta S(t) I(t)}{1+a I(t)}+\lambda R(t)\right) \mathrm{d} t+\sigma_{1}\left(S(t)-S^{*}\right) \mathrm{d} B(t), \\
&\mathrm{d} I(t)=\left(\frac{\beta S(t) I(t)}{1+\alpha I(t)}-(\mu+\gamma) I(t)\right) \mathrm{d} t+\sigma_{2}\left(I(t)-I^{*}\right) \mathrm{d} B(t), \\
&\mathrm{d} R(t)=(\gamma I(t)-(\mu+\lambda) R(t)) \mathrm{d} t+\sigma_{3}\left(R(t)-R^{*}\right) \mathrm{d} B(t),\\
&\text{+ positive initial conditions.}
\end{aligned}
\end{cases}
\end{equation}
With a natural birth rate $\Lambda >0,$ a natural death rate $\mu>0$ and different stochastic intensities $\sigma_1, \sigma_2 , \sigma_3 >0,$ and such that $(S^*,I^*,R^*)$ is the EE of the deterministic counterpart of the model obtained by taking $\sigma_1=\sigma_2=\sigma_3=0.$\\
The author established the mathematical well posedness of the model as well as the biological one. The stochastic permanence of the disease was proven to occur if 
$\max \left\{\sigma_{1}^{2}, \sigma_{2}^{2}, \sigma_{3}^{2}\right\}<2(\Lambda-\mu)$. Numerical simulations using the Milstein's Higher Order
Method \cite{sto} have been done for different values of the intensities to show the contribution of the additional noise to the model.  \\\\
\hspace*{0.3cm}  \cite{elkoufi} tried to adapt a deterministic SIR model with vaccination and vertical transmission to a stochastic point of view with a more generalized incidence function. Their proposed model was 
\begin{equation}
\begin{cases}
\begin{aligned}
&\hspace*{-0.2cm}d S(t)=\bigg(-\frac{\beta S(t) I(t)}{1+\alpha_{1} S(t)+\alpha_{2} I(t)+\alpha_{3} S(t) I(t)}-b S(t)+(1-m) p d I(t)\\
&\hspace*{0.8cm}+b(1-m)(S(t)+R(t))\bigg) d t-\frac{\sigma S(t) I(t)}{1+\alpha_{1} S(t)+\alpha_{2} I(t)+\alpha_{3} S(t) I(t)} d B(t), \\
&\hspace*{-0.2cm}d I(t)=\left(\frac{\beta S(t) I(t)}{1+\alpha_{1} S(t)+\alpha_{2} I(t)+\alpha_{3} S(t) I(t)}-(p d+\gamma) I(t)\right) d t\\
&\hspace*{0.7cm}+\frac{\sigma S(t) I(t)}{1+\alpha_{1} S(t)+\alpha_{2} I(t)+\alpha_{3} S(t) I(t)} d B(t), \\
&\hspace*{-0.2cm}d R(t)=(\gamma I(t)-b R(t)+d m p I(t)+m b(S(t)+R(t))) d t, \\
+&\text{ positive initial conditions.}
\end{aligned}
\end{cases}
\end{equation}
With the additional parameters $b>0$ (resp $d>0$) is the mortality rate in the susceptible and the
recovered (resp infective) individuals, $p$ is the proportion of the offspring of the infective and $q>0$ is the proportion of the rest that are born infected, with $p+q=1$ and $\alpha_1,\alpha_2,\alpha_3>0.$ Finally $m \in [0,1]$ is the successful vaccination proportion to the newborn from the susceptible and the recovered. \\
The value of the work presented by the authors is that it implicitly deals with all other known transmition rates (Beddington-DeAngellis, Crowley Martin, Holling-type) discussed earlier in this paper.\\
\hspace*{0.3cm} The authors proved the existence and uniqueness of a global positive solution to the model, they also proved that if either $\sigma^{2}>\dfrac{\beta^{2}}{(2(p d+r))}$ or ( $R_{s}<1$ and $\sigma^{2}<\beta$ ) the disease dies out, while if  $R^{*}_{s}>1,$ then the disease persists in the mean. Where $R_s$ (resp $R^{*}_s$) is the extinction (resp persistence) threshold given by 
$$ \begin{cases} 
R_{s}=\dfrac{R_{0}}{(1-m)}-\dfrac{\sigma^{2}}{2\left(1+\alpha_{1}(1-m)\right)^{2}(p d+r)}. \\ 
R_{s}^{*}=(1-m) R_{0}-\dfrac{\sigma^{2}}{2\left(1+\alpha_{1}(1-m)\right)^{2}(1-m)(p d+r)}.
\end{cases}
$$ such that $\mathcal{R}_0$ is the basic reproduction number associated to the deterministic counterpart of the proposed model (i.e for $\sigma=0$). The numerical simulation of the model confirmed the theoretical results, and a comparison of solutions with the deterministic case $\sigma=0$ was illustrated.
\subsection{Some of the main contributions to  \eqref{eq:fde}}
\hspace{0.3cm} The purpose of this type of contributions is to deal with \textbf{Limitation 2} by including the "memory effect" to the dynamics of the and adapting the previously introduced models and the contributions made as well. This is done by replacing the classical derivative with a well chosen fractional one. \\
\hspace*{0.3cm}  \cite{ozalp} proposed the following SEIR model
\begin{equation}
\begin{cases}
\begin{aligned}
&D^{\alpha} S(t)=\Lambda-\frac{p \Lambda E(t)}{N(t)}-\frac{q \Lambda I(t)}{N(t)}-r \frac{S(t) I(t)}{N(t)}-\mu S(t), \\
&D^{\alpha} E(t)=\frac{p \Lambda E(t)}{N(t)}+\frac{q \Lambda I(t)}{N(t)}+r \frac{S(t) I(t)}{N}-\mu E(t)-\beta E(t), \\
&D^{\alpha} I(t)=\beta E(t)-\mu I(t)-d I(t)-\gamma I(t), \\
&D^{\alpha} R(t)=\gamma I(t)-\mu R(t),\\
&\text{+positive initial conditions.}
\end{aligned}
\end{cases}
\end{equation}
such that $r>0$ is the horizontal transmission rate between the susceptible and the exposed. $p \in [0,1]$ resp ($q \in [0,1]$) is the probability of the offspring of exposed (resp infected) being born in the exposed class, $\beta>0$ is the rate of exposed individuals becoming infectious. and $D^\alpha$ is the Caputo fractionnal order derivative of order $\alpha \in (0,1)$ \cite{review}.\\

\hspace{0.3cm} The authors proved the mathematical well posedness of the model as well as the positiveness and boundedness of the unique solution. Under the assumption $\mathcal{R}_0:=\dfrac{(q \mu+r) \beta}{(\mu+d+\gamma)(\mu-p \mu+\beta)}<1$, the DFE was proven to be asymptotically stable. For the endemic equilibrium, the application of the fractional Routh-Hurwitz criterion (see \cite{rh1}) permitted the authors to obtain assumptions on the model's parameters under which the EE is asymptotically stable. The numerical simulation was carried on by means of the generalized Adam-Bashford method and an adequate choice of the parameters illustrated the theoretical stability results.\\\\
\hspace*{0.3cm} In the same year, to model Influenza,  \cite{el-shahed} replaced the classical derivative of the model proposed by \cite{adap} with the Caputo fractional one to get the following modified SIRC model
\begin{equation}
\begin{cases}
\begin{aligned}
&D^{\alpha} S(t)=\mu(1-S(t))-\beta S(t) I(t)+\gamma C(t), \\
&D^{\alpha} I(t)=\beta S(t) I(t)+\sigma \beta C(t) I(t)-(\mu+\theta) I(t), \\
&D^{\alpha} R(t)=(1-\sigma) \beta C(t) I(t)+\theta I(t)-(\mu+\delta) R(t), \\
&D^{\alpha} C(t)=\delta R(t)-\beta C(t) I(t)-(\mu+\gamma) C(t),\\
&\text{+ positive initial conditions.}
\end{aligned}
\end{cases}
\end{equation}
Such that the birth and death dates are equal to $\mu>0$, $\gamma^{-1}$ is the cross-immune period $\theta^{-1}$ is the infectious period, $\delta^{-1}$ is the total immune period and finally $\sigma$ is the fraction of the exposed cross-immune recruited into the infective subpopulation.\\\\
\hspace*{0.3cm} All the usual steps of study were followed, the existence, uniqueness,  boundedness and positiveness of the solution were proven. The local asymptotic stability of the FDE is assured under the condition $R_{0}:=\dfrac{\beta}{(\mu+\theta)}<1$. For the endemic equilibrium, the authors uded the fractional Routh-Hurwitz criterion (see \cite{rh1}) which allowed them to obtain conditions under which the EE is locally asymptotically stable. \\\\
\hspace*{0.3cm}   \cite{mouaouine} considered the following  SIR model with fractional Caputo derivative.
\begin{equation}
\begin{cases}
\begin{aligned}
&D^{\alpha} S(t)=\Lambda-\mu S(t)-\frac{\beta S(t) I(t)}{1+\alpha_{1} S(t)+\alpha_{2} I+\alpha_{3} S(t) I(t)}, \\
&D^{\alpha} I(t)=\frac{\beta S(t) I(t)}{1+\alpha_{1} S(t)+\alpha_{2} I(t)+\alpha_{3} S(t) I(t)}-(\mu+d+\gamma) I(t), \\
&D^{\alpha} R(t)=\gamma I(t)-\mu R(t),\\
&\text{+ positive initial conditions.}
\end{aligned}.
\end{cases}
\end{equation}
With $\alpha_1,\alpha_2,\alpha_3 >0.$ Note that this model holds the value of studying several models that can be considered with each incidence rate by studying only one with a generalized incidence rate. \\
\hspace*{0.3cm} As always, the mathematical and biological well posedness were established. The local asymptotic stability of the DFE (resp EE) was proven under the condition $\mathcal{R}_{0}:=\dfrac{\beta \Lambda}{\left(\mu+\alpha_{1} \lambda\right)(\mu+d+r)}\leq 1$ (resp $\mathcal{R}_0>1$), while using the Routh-Hurwitz criterion (see \cite{rh1}) for the EE case. The global stability  of the FDE and EE was proven under the same previous conditions by means of the Lyapunov method. Numerical simulations were carried out by using an algorithm based on the fractional Euler's method (see \cite{odib}) and  supported the theoretical results obtained.\\\\
\hspace*{0.3cm} In the aim of adapting the contribution of the generalization of the incidence rate to the fractional case,   \cite{khan2} proposed the two following SIR models
\begin{center}
\begin{equation}\label{eq:caputo} 
\begin{cases}
\begin{aligned}
&D_{C}^{\alpha} S(t)=\Lambda-\mu S(t)-S(t) f(I(t))+\mu_{1} I(t)+\theta R(t) \\
&D_{C}^{\alpha} I(t)=S(t) f(I(t))-\left(\mu+\mu_{1}+\mu_{2}+d \right) I(t) \\
&D_{C}^{\alpha} R(t)=\mu_{2} I(t)-(\mu+\theta) R(t),\\
&\text{+ positive initial conditions}.
\end{aligned}
\end{cases}
\end{equation}
\begin{equation}\label{eq:abc}
\begin{cases}
\begin{aligned}
&D_{ABC}^{\alpha} S(t)=\Lambda-\mu S(t)-S(t) f(I(t))+\mu_{1} I(t)+\theta R(t) \\
&D_{ABC}^{\alpha} I(t)=S(t) f(I(t))-\left(\mu+\mu_{1}+\mu_{2}+d \right) I(t) \\
&D_{ABC}^{\alpha} R(t)=\mu_{2} I(t)-(\mu+\theta) R(t),\\
&\text{+ positive initial conditions}.
\end{aligned}
\end{cases}
\end{equation}
\end{center}
Such that $D^\alpha_{C}$ (resp $D^\alpha_{ABC}$) is the Caputo (resp Atangana-Baleanu-Caputo) fractional derivative, $\mu_1>0$ resp ($\mu_2>0$) is the rate of movement of the infected to the susceptible (resp recovered) class and $\theta>0$ is the rate in which the recovered population becomes susceptible. $f$ was supposed to be non-negative and
continuously differentiable in $\mathbb{R}^{*}_{+},$ locally lipschitz on $\mathbb{R}_+$ and satisfying
\begin{enumerate}
\item[$\bullet$] $f(0)=0$ and $f(I)>0\;\; \forall I>0.$
\item[$\bullet$] $\dfrac{f(I)}{I}$ is continuous and monotonically non-increasing $\forall I>0$ and \\ $\lim _{I \rightarrow 0^{+}} \dfrac{f(I)}{I}$  exists. 
\end{enumerate}
\begin{rem} Although this model was considered by the authors to be an SIR model, the susceptibility of the recovered population in real life is generally due to immunity-loss to the disease. Taken that into account, the model is rather an SIRS model. \end{rem}
\hspace*{0.3cm} For both models, The authors proved the mathematical and biological well posedness of the two models. The DFE  was proven to be asymptotically locally and globally stable under the condition $\mathcal{R}_{0}:=\dfrac{\Lambda}{\mu \left(\mu+\mu_{1}+\mu_{2}+d \right)} \dfrac{\partial f(0)}{\partial I}<1$, whereas for the EE, it was proven to be asymptotically globally stable under the condition $\mathcal{R}_0>1.$ To show the importance of the choice of the incidence functions, numerical simulations were done for all three of them (Bilinear, Holling type and Beddington-DeAngellis). Interpretations  in terms of the fractional order $\alpha$ were given, precisely, the decreasing (resp increasing) of the susceptible and recovered (resp the infected) for increased values of $\alpha.$ The numerical simulation also proved the feasibility of the Atangana-Baleanu-Caputo fractional derivative over the Caputo one.
\subsection{Some of the contributions to  \eqref{eq:dde1}}
\hspace*{0.3cm} For this kind of contributions, \textbf{Limitation 3} is dealt with by  adding of the notion of delay to the model to get a model falling into category \eqref{eq:dde1}.\\\\
\hspace*{0.3cm}   \cite{kaddar} considered the following SIR model with a Beddington-DeAngelis incidence rate 
\begin{equation}
\begin{cases}
\begin{aligned}
&S'(t)=\Lambda-\mu S(t)-\frac{\beta S(t-\tau) I(t-\tau)}{1+\alpha_{1} S(t-\tau)+\alpha_{2} I(t-\tau)}, \\
&I'(t)=\frac{\beta S(t) I(t)}{1+\alpha_{1} S(t)+\alpha_{2} I(t)}-(\mu+d+\gamma) I(t), \\
&R'(t)=\gamma I(t)-\mu R(t),\\
&\text{+ positive initial functions.}
\end{aligned}
\end{cases}
\end{equation}
With $\alpha_1,\alpha_2>0$ and $\tau>0$ representing the incubation period.\\\\
\hspace*{0.3cm} The author proved that the FDE is locally asymptotically stable under the condition $\mathcal{R}_{0}:=\dfrac{\Lambda\left(\beta-\alpha_{1}(\mu+\alpha+\gamma)\right)}{\mu(\mu+\alpha+\gamma)}<1.$ One can remark that the incubation period does not affect the stability of the FDE. For the EE, under the assumptions $\mathcal{R}_0<1$ and $\alpha_{2} \mu<\beta-\alpha_{1}(\mu+\alpha+\gamma)$,  the author established the existence of a critical incubation period $\tau_0$ such that if $\tau \in [0,\tau_0)$ the EE is locally asymptotically stable and unstable  for $\tau>\tau_0.$ For the critical case $\tau=\tau_0$ bifurcation from the EE occurs. Numerical simulation was done separately for the three possible values of $\tau$ and confirmed the obtained theoretical results.\\\\
\hspace*{0.3cm} Global stability of an SIR model with Holling type incidence rate was studied by \cite{mccluskey2}. The considered model was 
\begin{equation}
\begin{cases}
\begin{aligned}
&S'(t)=\Lambda-\mu_1 S(t)-\frac{\beta S(t-\tau) I(t-\tau)}{1+a I(t-\tau)}, \\
&I'(t)=\frac{\beta S(t-\tau) I(t-\tau)}{1+a I(t)}-(\mu_2+\gamma) I(t), \\
&R'(t)=\gamma I(t)-\mu_3 R(t),\\
&\text{+ positive initial functions.}
\end{aligned}
\end{cases}
\end{equation}
Such that $\tau>0$ is the latent period and $\mu_1,\mu_2,\mu_3$ are respectively the natural death rates of the susceptible, infected and recovered population. For biological reasons, it was supposed that the removal of infectives is at least as fast as the removal of susceptibles the assumption. Mathematically speaking, $\mu_{1} \leq \mu_{2}+\gamma$.\\\\
\hspace*{0.3cm} It should be known that the model was studied earlier by  \cite{xu} in which they proved that the FDE is globally asymptotically stable under the condition $\mathcal{R}_{0}:=\dfrac{\Lambda \beta}{\mu_{1}\left(\mu_{2}+\gamma\right)}<1$ . Whereas, if $\mathcal{R}_0 >1$ and $\underbrace{\Lambda a>\mu_{2}+\gamma}_{(H)}$,  the EE is locally stable. McCluskey's contribution was discarding $(H)$ by considering the function $g(x):=x-1-ln(x)$ and using it to define the following carefully chosen Lyapunov function 
$$
V(t):=\frac{1}{\beta f\left(I^{*}\right)} g\left(\frac{S(t)}{S^{*}}\right)+\frac{I^{*}}{\beta S^{*} f\left(I^{*}\right)} g\left(\frac{I(t)}{I^{*}}\right)+\int_{0}^{\tau} g\left(\frac{I(t-s)}{I^{*}}\right) \mathrm{d}s
$$
where
$$
f(x)=\dfrac{x}{1+\alpha x}, \quad \alpha>0.
$$
\hspace*{0.3cm} In the aim of adapting the generalization of the incidence rate to this type of models, \cite{li2} proposed in 2014 the following model with a general incidence and recovery rates.
$$
\begin{cases}
\begin{aligned}
&S'(t)=\Lambda-\mu S(t)-F(S(t), I(t)) \\
&I'(t)=e^{-\mu \tau} F(S(t-\tau), I(t-\tau))-(\mu+d) I(t)-G(I(t)), \\
&R'(t)=G(I(t))-\mu R(t),\\
&\text{+ positive initial functions.}
\end{aligned}
\end{cases}
$$ 
Such that $\tau \geq 0$ is the latent
period and the term $e^{-\mu \tau}$ is  used to model the survival rate of the population. All along the authors' study the total assumptions fixed on the functions $F$ and $G$ are 
\begin{small}
\begin{enumerate}
\item $F(S,I)>0, \dfrac{\partial F(S,I)}{\partial I}>0, \dfrac{\partial F(S,I)}{\partial S}>0$ \;\; $\forall$ $S,I>0$.
\item $F(S, 0)=F(0, I)=0, \dfrac{\partial F(S,I)}{\partial S}\bigg|_{(S, 0)}=0, \dfrac{\partial F(S,I)}{\partial I}\bigg|_{(S, 0)}>0$ \;  $\forall S,I>0$.
\item $G(0)=0,\; G^{\prime}(I)>0$ \; $\forall I \geq 0.$
\item $F'(S,0)$ is increasing $\forall S>0$.
\item $F(S,I) \leq I \dfrac{\partial F(S,I)}{\partial I}\bigg|_{(S,0)} \forall I>0.$
\item $G^{\prime}(0) \leq \dfrac{G(I)}{I} \; \forall I>0.$
\item $\dfrac{I}{I^{*}} \leq \dfrac{F(S, I)}{F\left(S, I^{*}\right)}$  $\forall I \in\left(0, I^{*}\right), \dfrac{F(S, I)} {F\left(S, I^{*}\right)} \leq \dfrac{I}{ I^{*}}$ $\forall I \geq I^{*}.$
\item $\dfrac{G(I)}{G\left(I^{*}\right)} \leq \dfrac{I}{I^{*}}$ $\forall I \in\left(0, I^{*}\right)$ and $\dfrac{I}{ I^{*}} \leq \dfrac{G(I)}{G\left(I^{*}\right)}$  $\forall I \geq I^{*}$.
\end{enumerate}
\end{small}
\hspace*{0.3cm} The first condition of assumption 2  allowed the authors to prove the existence of the FDE. $\forall \tau>0,$ the assumptions $1,2$ and $3$ along with $\mathcal{R}_0:=\dfrac{e^{-\mu \tau}  \dfrac{\partial F(S,I)}{\partial I}\bigg|_{FDE}}{\mu+\alpha+G^{\prime}(0)}<1$ allowed them to prove the existence of the EE. For $\mathcal{R}_0\leq1,$  the local  asymptotic stability of the FDE  was established under the assumptions $1,2,3$ while it was disproven if $\mathcal{R}_0>1.$ For $\mathcal{R}_0 \leq 1,$ the global stability of the FDE was proven and required the assumptions $1-6.$ Whereas for $\mathcal{R}_0>1$ the assumptions $1,2$ and $3$ (resp $1,2,3$ and $7$)  were used to prove the existence (resp global asymptotic stability) of the EE. 
\begin{rem} While the term $e^{- \mu \tau}$ only shows in $\mathcal{R}_0.$ Comparing these results to the ones obtained by Kaddar in the model introduced in the beginning of the section by taking $F(S,I)=\dfrac{\beta S I}{1+\alpha_1 S+\alpha_2 I}$ and $G(I)=\gamma I,$ we deduce that no critical latent period showed in the analysis of \cite{li2} and the FDE was proven to be globally asymptotically stable only under the assumption $\mathcal{R}_0>1$. Implying that the survival of the population plays an important role in the dynamics of the disease.\end{rem}
\subsection{Some of the main contributions to  \eqref{eq:pde}}
\hspace*{0.3cm} For this category of contribution, the aim is to deal with \textbf{Limitation 4} by introducing the spacial factor to the considered models. \\
 \cite{settapat} proposed the following  SIR model 
\begin{equation}\label{eq:prev}
\begin{cases}
\begin{aligned}
&\dfrac{\partial S(x,t)}{\partial t}-\alpha \Delta S(x,t)=\mu N(x,t)-\mu S(x,t)-\beta S(x,t) I(x,t), & x \in \Omega, t>0, \\
&\dfrac{\partial I(x,t)}{\partial t}-\alpha \Delta I(x,t)=-(\mu+\gamma) I(x,t)+\beta S(x,t) I(x,t), & x \in \Omega, t>0, \\
&\dfrac{\partial R(x,t)}{\partial t}-\alpha \Delta R(x,t)=\gamma I+\mu R(x,t), & x \in \Omega, t>0,\\
&\partial_{\eta} S(x,t) =\partial_{\eta} I(x,t)=\partial_{\eta} R(x,t)=0, & x \in \partial \Omega, t>0,\\
&\text{+ positive initial conditions.}
\end{aligned}
\end{cases}
\end{equation}
With birth and death rates equal to $\mu>0,$ and $\alpha>0$ being the rate of propagation of the individuals.\\
\hspace*{0.3cm} The authors proved the existence and uniqueness of the solution for a maximum time of existence $T_{max}.$ From the boundedness of the solution it was deduced that $T_{max}=\infty.$ The positiveness of the solution was an immediate consequence of the maximum principle. The local asymptotic stability of the DFE was proven (resp disproven) under the assumption $\mathcal{R}_{0}:=\dfrac{N \beta}{\mu+\gamma}<1$ (resp $\mathcal{R}_0 > 1$), the global asymptotic stability was carried out by the Lyapunov method. The EE was proven to be locally (resp globally) asymptotically stable under the condition $\mathcal{R}_0>1.$ The numerical simulation was restricted to the one dimensional case in space. The authors proposed a numerical scheme based on the  forward approximation on time and a $\theta$ approximation in space. The numerical results supported the theoretical ones, and the infected population was proven to spread more in space in terms of increased values of the diffusion rate $\alpha$.  
\begin{rem}The conditions $\partial_\eta S(x,t)=\partial_\eta I(x,t)=\partial_\eta R(x,t)=0$ model the confinement of the population, on the other hand, while the diffusion rates are taken in this model to be constant, for contagious diseases, they can depend on the infected population. When the infected increase, individuals tend not to propagate in space to not get infected. It would be interesting to restudy the model by considering infected depending diffusion rates. One last remark concerns the biological feasibility of the proposed scheme, although not done by the authors, the numerical scheme proposed is proven to  be positive invariant, that is, the iterates of the approximate solution are always positive. \end{rem} 
\hspace*{0.3cm} In the aim of contribution to the generalization of the incidence rate to models falling into category $\eqref{eq:pde},$ \cite{lotfi} proposed the following generalized diffusive SIR model
\begin{equation}
\begin{cases}
\begin{aligned}
&\dfrac{\partial S(x,t)}{\partial t}=d_{S} \Delta S(x,t)+\Lambda-\mu S(x, t)\\ 
&\hspace*{1.5cm}-\dfrac{\beta S(x, t) I(x, t)}{1+\alpha_{1} S(x, t)+\alpha_{2} I(x, t)+\alpha_{3} S(x, t) I(x, t)} \; x \in  \Omega, t>0,\\
&\dfrac{\partial I(x,t)}{\partial t}=d_{I} \Delta I(x,t)-(\mu+d+\gamma) I(x, t)\\
&\hspace*{1.5cm} +\frac{\beta S(x, t) I(x, t)}{1+\alpha_{1} S(x, t)+\alpha_{2} I(x, t)+\alpha_{3} S(x, t) I(x, t)}  \; x \in  \Omega, t>0,
 \\
&\dfrac{\partial R(x,t)}{\partial t}=d_{R} \Delta R(x,t)+\gamma I(x, t)-\mu R(x, t) \; x \in \Omega, t>0,\\
&\partial_{\eta} S(x,t)=\partial_{\eta} I(x,t)=\partial_{\eta} R(x,t)=0, \; x \in \partial \Omega, t>0,\\
&\text{+ positive initial conditions.}
\end{aligned}
\end{cases}
\end{equation}
with $d_S,d_I>0$ and $d_R>0$ being the diffusion rates associated to the susceptible, infected and recovered population respectively while $\alpha_1,\alpha_2,\alpha_3 >0.$\\\\
By the same analogy of the previous model \eqref{eq:prev}, the authors proved the existence and uniqueness of a global, positive and bounded solution to the model. Knowing that $\mathcal{R}_{0}:=\dfrac{\beta \Lambda}{\left(\mu+\alpha_{1} \Lambda\right)(\mu+d+\gamma)},$ a linearization of the system around the FDE (resp EE) proved the latter to be asymptotically locally stable under the condition $\mathcal{R}_0<1$ (resp $\mathcal{R}_0>1$)  while it was disproven for the FDE in the case $\mathcal{R}_0>1.$ Inspired by the Lyapunov function given by \cite{hattaf}, the authors were able to prove the global asymptotic stability of the FDE (resp EE) under the condition $\mathcal{R}_0\leq1$ (resp $\mathcal{R}_0>1$). The numerical simulation supported the theoretical stability results, moreover, as in \eqref{eq:prev}, for the case $\mathcal{R}_0>1$ and a high diffusion rate $d_I,$ the population of the infected spreads in space.\\\\
\hspace*{0.3cm} Knowing that the diffusion plays an important role, the question of controlling the disease in the case of fast spread arises. Taking that into account,  \cite{adnaoui} considered the same previous model but were interested into studying the existence of a spatiotemporal optimal control representing the vaccination. The modified model was expressed by
\begin{equation}
\begin{cases}
\begin{aligned}
&\dfrac{\partial S(x,t)}{\partial t}=d_{S} \Delta S(x,t)+\Lambda-(\mu+v(x,t)) S(x, t) \\
&\hspace*{1.5cm}-\dfrac{\beta S(x, t) I(x, t)}{1+\alpha_{1} S(x, t)+\alpha_{2} I(x, t)+\alpha_{3} S(x, t) I(x, t)} \; x \in  \Omega, t>0,\\
&\dfrac{\partial I(x,t)}{\partial t}=d_{I} \Delta I(x,t)-(\mu+d+\gamma) I(x, t) \\
&\hspace*{1.5cm} +\frac{\beta S(x, t) I(x, t)}{1+\alpha_{1} S(x, t)+\alpha_{2} I(x, t)+\alpha_{3} S(x, t) I(x, t)} \; x \in  \Omega, t>0, \\
&\dfrac{\partial R(x,t)}{\partial t}=d_{R} \Delta R(x,t)+\gamma I(x, t)-\mu R(x, t)+v(x,t)S(x,t) \; x \in \Omega, t>0,\\
&\partial_{\eta} S=\partial_{\eta} I=\partial_{\eta} R=0, \; x \in \partial \Omega, t>0,\\
&\text{+ positive initial conditions.}
\end{aligned}
\end{cases}
\end{equation}
\hspace*{0.3cm} The authors proved the mathematical and biological well posedness of the model with vaccination which allowed to prove the existence of an optimal solution to the optimal control problem in which the aim is the minimization (over a biologically feasible set) of the following objective functional 
\begin{equation}
J(v)=\int_{0}^{T} \int_{\Omega}\left(\rho_{1} S(x, t)+\rho_{2} I(x, t)\right) d x d t+\frac{\alpha}{2}\|v\|_{L^{2}(\Omega \times [0,T])}^{2}.
\end{equation}
where $\alpha,\rho_1,\rho_2>0$ are  chosen weighted constants. The characterization of the optimal control $v^*$ was given in terms of the optimal state by the following formula $v^{*}=\min \left(v^{\max }, \max \left(0, \dfrac{y_{1}^{*} p_{1}-y_{1}^{*} p_{3}}{\alpha}\right)\right),$ where $p$ is the solution to the adjoint problem. The numerical simulation was carried on by the forward-backward sweep method in which the direct problem is solved forward in time, and the adjoint problem backwards. A choice of parameters corresponding to a situation in which the fast spread of the disease occurs was illustrated by simulating the model without control. In the case with control, interesting results have been deduced. Primarily, the major role of the immediate vaccination (1 day after the beginning of the disease) of the population in controlling the spread of the disease over the delayed one (20 days after).
\section{Conclusion}
\hspace*{0.3cm} The primal goal of this paper was to familiarize future researchers  with the existing types of compartmental models in epidemiology as  well as the contributions' scope of each 
type. The paper takes into account some of the main contributions in each category, but by no means covers them all, since that would require a whole book. The reader should know that the ongoing and future works are in the aim of developing more complex models combining the reviewed types $\eqref{eq:sde},\eqref{eq:fde},\eqref{eq:dde1}$ and $\eqref{eq:pde}.$ We refer for instance to the following recent works \cite{f1,f2,f3,f4}.\\\\ 
\hspace*{0.3cm} Knowing all these types, it can be concluded that, while modeling a disease, the pertinence of the chosen model  is indeed based on its capacity of maximizing the interpretation of reality but also on its ability to minimize the difficulty of the required mathematical and numerical tools for its analysis. And the level that one can sacrifice on each side is based on the better understanding of the modeled disease, such understanding can be gained in terms of contributions with experts in the modeled field, in this case, biologists. 
\bibliographystyle{agsm}
\begin{spacing}{0.2}
\begin{small}
\bibliography{bibliography}
\end{small}
\end{spacing}
\end{document}